\documentstyle[12pt,psfig]{article}

\catcode`\@=11
\long\def\@makefntext#1{
\protect\noindent \hbox to 3.2pt {\hskip-.9pt
$^{{\ninerm\@thefnmark}}$\hfil}#1\hfill}                

\def\@makefnmark{\hbox to 0pt{$^{\@thefnmark}$\hss}}  

\def\ps@myheadings{\let\@mkboth\@gobbletwo
\def\@oddhead{\hbox{}
\rightmark\hfil\ninerm\thepage}
\def\@oddfoot{}\def\@evenhead{\ninerm\thepage\hfil
\leftmark\hbox{}}\def\@evenfoot{}
\def\sectionmark##1{}\def\subsectionmark##1{}}

\renewcommand{\thefootnote}{\fnsymbol{footnote}}

\def\sectionc{\@startsection {section}{1}{\z@}{-3.5ex plus -1ex minus
    -.2ex}{2.3ex plus .2ex}{\bf }}
\def\subsectionc{\@startsection{subsection}{2}{\z@}{-3.25ex plus -1ex minus
   -.2ex}{1.5ex plus .2ex}{\it }}
\renewcommand{\section}[1]{\sectionc{#1}\hspace*{\parindent}}
\renewcommand{\subsection}[1]{\subsectionc{#1}\hspace*{\parindent}}
\newcounter{appendixc}
\newcounter{subappendixc}[appendixc]
\newcounter{subsubappendixc}[subappendixc]

\renewcommand{\appendix}[1] {\vspace*{0.6cm}
        \refstepcounter{appendixc}
        \setcounter{figure}{0}
        \setcounter{table}{0}
        \setcounter{equation}{0}
        \renewcommand{\thefigure}{\Alph{appendixc}.\arabic{figure}}
        \renewcommand{\thetable}{\Alph{appendixc}.\arabic{table}}
        \renewcommand{\theappendixc}{\Alph{appendixc}}
        \renewcommand{\theequation}{\Alph{appendixc}.\arabic{equation}}
        \noindent{\bf Appendix \theappendixc #1}\par\vspace*{0.4cm}}

\def\abstracts#1{{

\centering{\begin{minipage}{13.2truecm}\footnotesize
\baselineskip=13pt\noindent
        \parindent=0pt #1
        \end{minipage}}\par}}

\renewenvironment{thebibliography}[1]
        {\begin{list}{\arabic{enumi}.}
        {\usecounter{enumi}\setlength{\parsep}{0pt}
\setlength{\leftmargin 0.75cm}{\rightmargin 0pt}
         \setlength{\itemsep}{0pt} \settowidth
        {\labelwidth}{#1.}\sloppy}}{\end{list}}

\newcounter{itemlistc}
\newcounter{romanlistc}
\newcounter{alphlistc}
\newcounter{arabiclistc}

\newcommand{\fcaption}[1]{
        \refstepcounter{figure}
        \setbox\@tempboxa = \hbox{\footnotesize Figure~\thefigure. #1}
        \ifdim \wd\@tempboxa > 6in
           {\begin{center}
        \parbox{6in}{\footnotesize\baselineskip=13pt Figure~\thefigure. #1}
            \end{center}}
        \else
             {\begin{center}
             {\footnotesize Figure~\thefigure. #1}
              \end{center}}
        \fi}

\newcommand{\tcaption}[1]{
        \refstepcounter{table}
        \setbox\@tempboxa = \hbox{\footnotesize Table~\thetable. #1}
        \ifdim \wd\@tempboxa > 6in
           {\begin{center}
        \parbox{6in}{\footnotesize\baselineskip=13pt Table~\thetable. #1}
            \end{center}}
        \else
             {\begin{center}
             {\footnotesize Table~\thetable. #1}
              \end{center}}
        \fi}

\def\@citex[#1]#2{\if@filesw\immediate\write\@auxout
        {\string\citation{#2}}\fi
\def\@citea{}\@cite{\@for\@citeb:=#2\do
        {\@citea\def\@citea{,}\@ifundefined
        {b@\@citeb}{{\bf ?}\@warning
        {Citation `\@citeb' on page \thepage \space undefined}}
        {\csname b@\@citeb\endcsname}}}{#1}}

\newif\if@cghi
\def\cite{\@cghitrue\@ifnextchar [{\@tempswatrue
        \@citex}{\@tempswafalse\@citex[]}}
\def\citelow{\@cghifalse\@ifnextchar [{\@tempswatrue
        \@citex}{\@tempswafalse\@citex[]}}
\def\@cite#1#2{{$\null^{#1}$\if@tempswa\typeout
        {IJCGA warning: optional citation argument
        ignored: `#2'} \fi}}

 1
 1
 1

\font\ninerm=cmr9

\begin{document}

\hfill KFA-IKP(Th)-1996-13

\bigskip\bigskip

\centerline{\normalsize\bf THE ROLE OF MASSIVE STATES IN}
\baselineskip=15pt
\centerline{\normalsize\bf 
CHIRAL PERTURBATION THEORY\footnote{Invited talk, CEBAF/INT N*
workshop, Seattle, September 1996}}

\vspace*{0.6cm}
\centerline{\footnotesize ULF-G. MEI{\ss}NER}
\baselineskip=13pt
\centerline{\footnotesize\it Forschungszentrum J\"ulich, Inst. f.
Kernphysik (Theorie)}
\baselineskip=13pt
\centerline{\footnotesize\it D-52425 J\"ulich, Germany}
\centerline{\footnotesize E-mail: Ulf--G.Meissner@kfa-juelich.de}
\vspace*{0.3cm}

\vspace*{0.6cm}
\abstracts{I review some basic facts about the chiral limit of QCD.
This allows to formulate an effective field theory below the chiral
symmetry breaking scale, chiral perturbation theory (CHPT). I show
that for threshold reactions, the spectrum of QCD is most economically
encoded in a set of coupling constants of operators of higher chiral
dimension. A consistent scheme to incorporate the $\Delta(1232)$ is
also discussed and some examples are given. It is stressed that more
precise low--energy data are needed to further test and sharpen 
the resonance saturation hypothesis in the presence of baryons.}

\normalsize\baselineskip=15pt
\setcounter{footnote}{0}
\renewcommand{\thefootnote}{\alph{footnote}}

\section{Effective field theory of QCD}\label{sec:EFT}
\noindent In the sector of the three light quarks, one can write
the QCD Lagrangian as 
\begin{equation}
{\cal L}_{\rm QCD} = {\cal L}_{\rm QCD}^0 - \bar{q}\, {\cal M} \, q
\, \, , \label{lqcd}
\end{equation}
with $q^T = (u,d,s)$ and ${\cal M} = {\rm diag}(m_u , m_d , m_s)$ the
current quark mass matrix. The current quark masses are believed to be
small compared to the typical hadronic scale, $\Lambda_\chi \simeq 1$~GeV.
${\cal L}_{\rm QCD}^0$ admits a global chiral symmetry, i.e. one can
independently rotate the left-- and right--handed components of the
quark fields. This symmetry is spontaneoulsy broken down to its
vectorial subgroup, SU(3)$_{L+R}$, with the appeareance of eight
massless Goldstone bosons. The explicit chiral symmetry breaking due
to the quark mass term gives these particles, identified with the
pions, kaons and eta (denoted $\phi$), a small mass. The consequences of the
spontaneous and the explicit chiral symmetry breaking can be
calculated by means of an effective field theory (EFT), called
chiral perturbation theory.\cite{wein,gl85} ${\cal L}_{\rm QCD}$ is
mapped onto an effective Lagrangian with hadronic degrees of freedom,
\begin{equation}
{\cal L}_{\rm QCD} = {\cal L}_{\rm eff}[U, \partial U , \ldots,  {\cal
  M} , B]
\, \, , \label{leff}
\end{equation}
where the matrix--valued field $U(x)$ parametrizes the Goldstones,
${\cal M}$ keeps track of the explicit symmetry violation and $B$
denotes matter fields (like e.g. the baryon octet). While the latter are
not directly related to the symmetry breakdown, their interactions are
severely constrained by the non--linearly realized
chiral symmetry and one can thus incorporate
them unambiguously.\cite{cwz} ${\cal L}_{\rm eff}$ admits an energy
expansion,
\begin{equation}
{\cal L}_{\rm eff} = {\cal L}_{\phi}^{(2)} + {\cal L}_{\phi}^{(4)} 
+ {\cal L}_{\phi B}^{(1)}+ {\cal L}_{\phi B}^{(2)}+ {\cal L}_{\phi B}^{(3)}
+ {\cal L}_{\phi B}^{(4)} + \ldots\, \, , \label{leffex}
\end{equation}
where the superscript $(i)$ refers to the number of derivatives and or
meson mass insertions. 
The first two terms in Eq.\ref{leffex} comprise the meson
sector \cite{gl85} whereas the next four are relevant for processes
involving one single baryon. The ellipsis stands for terms with more
baryon fields and/or more derivatives. The various terms contributing
to a certain process are organized by their {\it chiral} dimension $D$
(which differs in general from the physical dimension) as follows:\cite{cnpp}
\begin{equation}
D = 2L + 1 \sum_d (d-2) N_d^{\phi} + \sum_d (d-1) N_d^{\phi B} \, \, ,
\label{chidim}
\end{equation}
with $L$ the number of (Goldstone boson) loops and $d$ the vertex dimension
(derivatives or factors of the pion mass). Lorentz invariance and
chiral symmetry demand that $d \ge 2$ ($\ge 1$) for mesonic
(pion--baryon) interactions. So to lowest order, one has to deal with
tree diagrams ($L=0$) which is equivalent to the time--honored current
algebra (CA). However, we are now in the position of {\it systematically}
calculating the corrections to the CA results. It is also important to
point out  that ${\cal L}_{\phi}^{(4)}$ and 
${\cal L}_{\phi B}^{(2,3,4)}$ contain parameters not fixed by symmetry,
the so--called {\it low-energy} {\it constants} (LECs). These have to be
determined from data or are estimated from resonance exchange,\cite{reso} 
as discussed in some detail below.  The whole machinery is well 
documented, see e.g. Refs.\cite{ulfrev},\cite{eckerr},\cite{bkmrev}

\section{Decoupling theorem}\label{sec:DT}
The chiral limit of QCD exhibits some interesting features. In
particular, certain quantities show a non--analytic behaviour in
terms of the quark masses. A well-known example is the 
lowest order Goldstone boson loop 
contribution to the baryon mass,
\begin{equation}
\delta m_B = {\rm const} \cdot M_\phi^3 \sim m_{\rm quark}^{3/2} \,\, .
\label{mN}
\end{equation}
The constant in front of the $M_\phi^3$ is given in terms of known
parameters. More important, if one would calculate the same one loop
diagram with massive (resonance) intermediate states, its value would
remain unchanged but some higher order corrections ($\sim M_\phi^4)$
would get modified. This is the essence of the {\it decoupling} 
{\it theorem} derived some 15 years ago by Gasser and Zepeda.\cite{GZ} 
It states that the {\it leading} non--analytic corrections (LNAC) to
S--matrix elements and transition currents are given solely in terms
of ${\cal L}_{\rm eff} [U,B]$ (in the chiral limit), i.e. that 
the inclusion of
mesonic ($\rho, \omega, \ldots$) or baryonic resonances ($\Delta,
N^\star, \ldots$) in the pertinent loop graphs does not modify the
LNACs. Note also that tree graphs are obviously analytic in the quark
masses. A detailed exposition of these results is given in Ref.\cite{GZ} 
Another such effect is observed in the electromagnetic
polarizabilities of the proton. To lowest order in the chiral
expansion, these are given by one loop graphs and thus only depend on 
known parameters,\cite{bkmpola}
\begin{equation}
\bar{\alpha}_p = 10 \, \bar{\beta}_p = \frac{5e^2 g_A^2}{384 \pi^2 F_\pi}
\frac{1}{M_\pi} = \frac{5Cg_A^2}{4 M_\pi} =
12.4 \cdot 10^{-4} \, {\rm fm}^3 \,\, ,
\label{pola}
\end{equation}
which scales like $1/ \sqrt{m_{\rm quark}}$ and is in good agreement
with the empirical numbers, $\bar{\alpha}_p = (12.1 \pm 0.8 \pm 0.5) \cdot
10^{-4}\,$fm$^3$, $\bar{\beta}_p = (2.1 \mp 0.8 \mp 0.5) \cdot 10^{-4}
\,$fm$^3$.\cite{macg} However, it is well known that tree graphs
with an intermediate $\Delta (1232)$ contribute roughly $+10 \cdot 10^{-4}
\,$fm$^3$ to $\bar{\beta}_p$ and one thus might question the relevance
of the CHPT result. Here, the decoupling theorem and pion loops come
to ones rescue. As shown in Ref.,\cite{bkmpola2} the large
next-to-leading order tree $\Delta (1232)$ contribution to $\bar{\beta}_p$ is
almost completely cancelled by a pion loop graph at the same order
with a subleading non--analyticity $\sim \ln M_\pi$, which has a large 
positive coefficient,
\begin{equation}
\bar{\beta}_p = \frac{Cg_A^2}{8M_\pi} + \frac{C}{\pi} \biggl[
\Big( \frac{3(3+\kappa_s)g_A^2}{m}-c_2\Big)\ln\frac{M_\pi}{\lambda}
\biggr] + \delta \bar{\beta}_p^r (\lambda) \,\, ,
\end{equation}
where I have dropped a small finite piece from the loops. The value
for $c_2$ is given below and $\bar{\beta}_p^r$ subsumes the large 
$\Delta$--contribution from the dimension four counter terms of the
type ${\cal L}_{\pi N}^{(4)} \sim \bar{N} F_{\mu \nu} F^{\mu \nu} N$.
 Yet another neat example has been discussed by
Mallik in connection with terms of the type $M_\pi^4 \ln
M_\pi^2$ contributing to the nucleon mass.\cite{samir} 
 
\section{Resonance saturation}\label{sec:reso}
In the meson sector at next--to-leading order, the effective
Lagrangian ${\cal L}_{\phi}^{(4)}$ contains ten LECs, called $L_i$.
These have been determined from data in Ref.\cite{gl85} 
(for an update, see e.g. Ref.\cite{daphne}). The actual values of the
$L_i$ can be understood in terms of resonance exchange\cite{reso},
i.e. the renormalized $L_i^r (M_\rho)$ are practically saturated by 
resonance exchange ($S,P,V,A$). In some few cases, tensor mesons can
play a role.\cite{DT} This is sometimes called {\it chiral
  duality} because part of the excitation spectrum of QCD reveals itself
in the values of the LECs. Furthermore, whenever vector and axial
resonances can contribute, the $L_i^r (M_\rho)$ are completely
dominated by $V$ and $A$ exchange, called {\it chiral} {\it VMD}.\cite{DRV}
As an example, consider the finite (and thus scale--independent) LEC
$L_9$. Its empirical value is $L_9 = (7.1 \pm 0.3) \cdot 10^{-3}$. The
well--known $\rho$--meson (VMD) exchange model for the pion form
factor, $F_\pi^V (q^2) = M_\rho^2/ (M_\rho^2 -q^2)$ (neglecting the
width) leads to
$L_9 = F_\pi^2 /(2 M_\rho^2) = 7.2 \cdot  10^{-3}$,
in good agreement with the empirical value.
Even in the symmetry breaking sector related to the quark mass, where
only scalar and (non-Goldstone) pseudoscalar mesons can contribute,
resonance exchange helps to understand why SU(3) breaking is generally
of ${\cal O}(25\%)$, except for the Goldstone boson masses. In
principle, these LECs are calculable from QCD, a first attempt using
the lattice has been reported in Ref.\cite{RM} and these studies are 
continuing. 

\begin{figure}[bht]
\begin{center}
\vskip 1.0cm
\hskip -.5cm
\psfig{figure=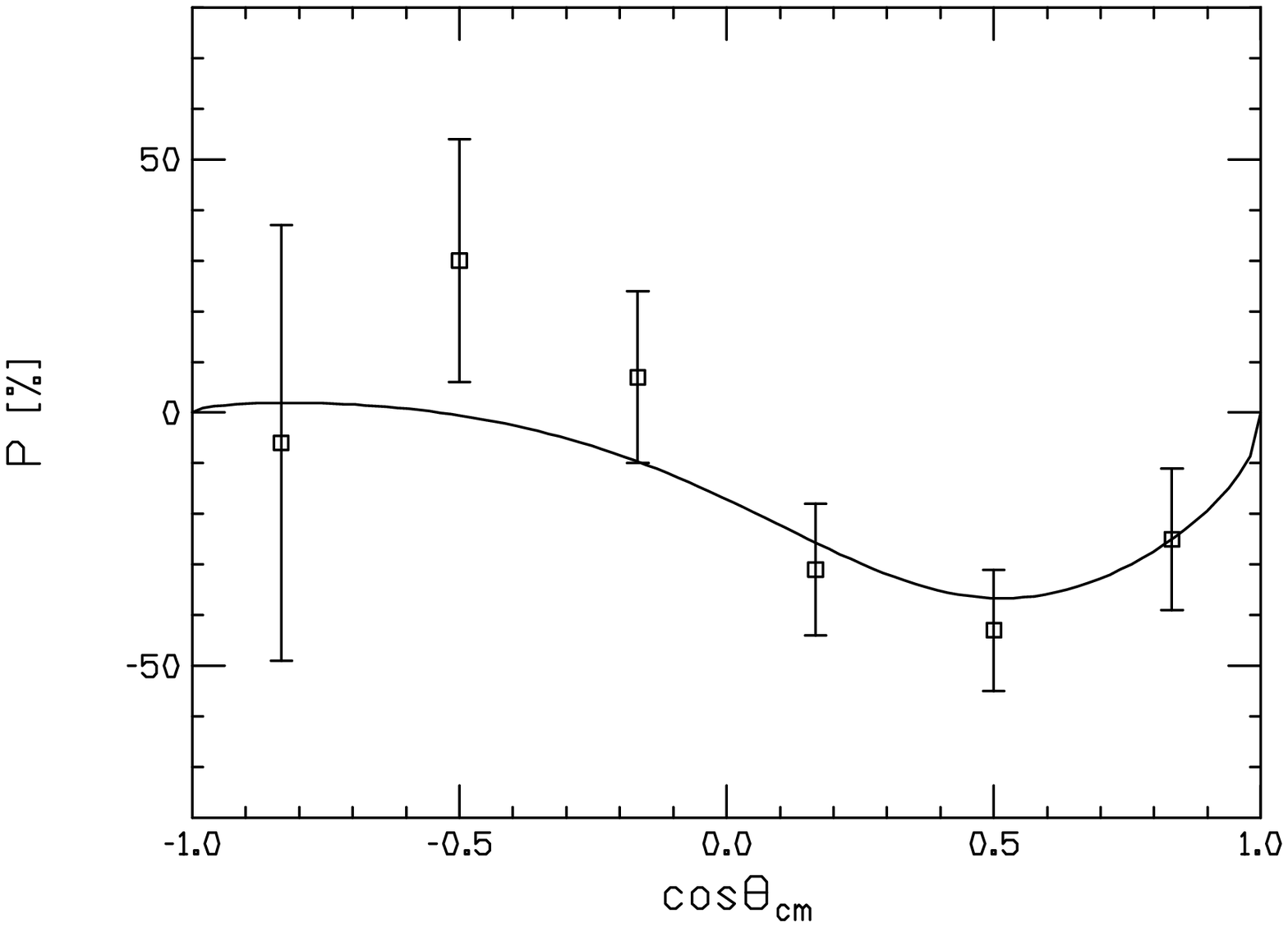,height=1.9in}%
\end{center}
\fcaption{Recoil polarization for $\gamma p \to K^+\Lambda$ calculated
in SU(3) HBCHPT to order $q^3$.
\label{fig:kaon}}
\end{figure}

Matters are much more complicated in the baryon sector, largely due to
the fact that one can have baryonic ($N^\star$) as well as mesonic ($M$)
excitations, symbolically 
\begin{equation}
\tilde{{\cal L}}_{\rm eff} [U,B;M,N^\star] \to {\cal L}_{\rm eff}
[U,B] \,\,\, , \label{Lreso}
\end{equation}
so that the LECs are given in terms of masses ($m_M, m_{N^\star}$) and
mesonic as well as baryonic coupling constants. Of the baryon
resonances, the $\Delta (1232)$ plays a particular role as explained 
below. Consider first an example where resonance saturation works
fairly well, namely the reaction $\gamma \, p \to \pi^0 \,p$ in the 
threshold region. The most accurate CHPT calculation performed so far
contains three LECs, two related to the S-wave $E_{0+}$ (called $a_1$
and $a_2$) and one related to the P-wave $P_3$, called 
$b_P$.\cite{bkmpi0},\cite{bkmpi02} Note that in the threshold region
one is only sensitive to the sum $a_1+a_2$ of the S--wave LECs.
A fit to the TAPS data\cite{fuchs}
leads to $a_1+a_2 = 6.60\,$GeV$^{-4}$ and $b_P=13.0\,$GeV$^{-3}$.
Note that the SAL data require a somewhat larger value for $b_p$.\cite{sal} 
Resonance exchange leads to  a completely fixed vector meson
contribution and to one from the $\Delta$, which depends on some
off--shell parameters. Constraining these from previous investigations
of the proton magnetic polarizability and the $\pi N$ P-wave scattering
volume $a_{33}$, one finds $a_1+a_2 = (a_1+a_2)^\Delta + (a_1+a_2)^V=
3.92+2.67 = 6.59\,$GeV$^{-4}$ together with $b_P^{\Delta +V} 
=13.0\,$GeV$^{-3}$ in very good agreement with the numbers obtained in
the free fit. 
Still, there remain
uncertainties to be clarified. Consider e.g. the four finite SU(2)
LECs $c_{1,2,3,4}$ related to the dimension two pion--nucleon
Lagrangian. These have been determined from one loop calculations of
the $\sigma$--term and certain $\pi N$ scattering observables 
to order $q^3$ but also
from the subthreshold expansion of the $\pi N$ scattering amplitude to
order $q^2$.\cite{bkmppn} The resulting values for the $c_i$ are
typically a factor 1.5 smaller than in Ref.\cite{bkmrev}.
We have recently reevaluated these coupling constants by constructing
observables which to one loop order $q^3$ are given entirely by tree
graphs with insertions from the dimension one and two Lagrangian and
have finite loop contributions, but are free of insertions from
${\cal L}_{\pi N}^{(3)}$.\cite{bkmlec} 
\renewcommand{\arraystretch}{1.3}
\begin{table}[hbt] 
\protect
\tcaption{Values of the dimension two LECs $c_i ' = 2mc_i$
$(i=1,\ldots,5)$ as determined in.\cite{bkmlec} 
The uncertainties on these parameters are discussed
in detail in that reference. The $^*$ denotes an input 
quantity.}\label{tab:ci}
\small
\vspace{0.4cm}
\begin{center}
\begin{tabular}{|c|c|c|r|r|}
\hline
\hline
       & ocuurs in & determined from & central value & res. exch. \\
\hline
$c_1'$ & $m_N,\sigma_{\pi N}$, $\gamma N \to \gamma N$ 
                          & phen. + res.exch. & $-1.7 $ & $-1.7^*$ \\
$c_2'$ & $\pi N \to \pi (\pi) N$, $\gamma N \to \gamma N$     
                          & phen. + res.exch. & $6.7 $ & $7.3$ \\
$c_3'$ & $\pi N \to \pi (\pi) N$, $\gamma N \to \gamma N$     
                          & phen. + res.exch. & $-10.1 $ & $-9.8$ \\
$c_4'$ & $\pi N \to \pi (\pi) N$    & phen. + res.exch. & $6.8 $ & $6.6$ \\
$c_5'$ & $(m_n-m_p)^{\rm strong}, \pi^0 N \to \pi^0 N $  
                                           & phen. & $-0.17 $ & --- \\
$c_6$ & $\kappa_p$, $\kappa_n$   & phen. + res.exch. & $5.8 $ & $6.1$ \\
$c_7$ & $\kappa_p$, $\kappa_n$   & phen. + res.exch. & $-3.0 $ & $-3.1$ \\
\hline
\hline
\end{tabular}
\end{center}
\end{table} 
The central values are listed in table~\ref{tab:ci}. These values can 
also be understood from resonance exchange assuming only that $c_1$ is
entirely saturated by scalar meson exchange. Amazingly, the ratio of
the scalar meson mass to scalar--meson--nucleon coupling needed,
$M_S / \sqrt{g_S} = 180\,$MeV, is exactly the one found in the Bonn
potential, where the scalar--isoscalar $\sigma$--meson models the
strong pionic correlations in the presence of nucleons ($M_\sigma = 550\,$MeV
and $g_\sigma^2/(4\pi)=7.1$).\cite{bonn}
Notice also that resonance saturation via vector meson exchange works
well for the anomalous magnetic moments. In the isoscalar case, we
find $\kappa_s = -0.12$
compared to $\kappa_\omega = -0.16$.\cite{mmd} For the isovector moment,
the empirical value $(\sim 5.8$) agrees well with the large 
tensor--to--vector coupling of the $\rho$, $\kappa_\rho \simeq 6$.
Also, Moj\v zi\v s has
recently analyzed other $\pi N$ scattering observables and finds somewhat
different values.\cite{martin} This remains to be clarified. In the
three--flavor sector, the LECs related to the symmetry breaking terms in the
meson--baryon Lagrangian can not yet simply be understood in terms of
scalar meson exchange. One would either need unphysically large scalar--baryon
couplings or low scalar meson masses (for details, see Ref.\cite{bora}).
At present, however, one has to use resonance saturation in actual
calculations since too few accurate low--energy data exist to pin down
all LECs (or at least a subset for $\pi N$ scattering and photo
reactions). An example where resonance saturation seems to work even for SU(3)
is shown in Fig.~\ref{fig:kaon}.\cite{sven} 
In that reference, kaon photo-- and
electroproduction off protons is considered. A few of the 13 LECs could be
fixed from single nucleon properties, the larger number from resonance
exchange.  
Amazingly, the shape of the recoil polarization for $\gamma p
\to K^+\Lambda$ measured at ELSA\cite{elsa} is well described (even so
the energy at which it was measured is somewhat too high).
It is important to stress that for threshold reactions,
the resonances need not be taken as dynamical degrees of freedom in the
effective field theory but rather should be used to estimate the
coefficients of higher dimension operators. However, if one is interested
in extending the range of applicability of CHPT say in the region of
the $\Delta (1232)$, it has to be accounted for dynamically.

\section{Above threshold: Inclusion of the $\Delta(1232)$}\label{sec:delta}
Although the inclusion of the decuplet was originally formulated for
SU(3),\cite{jmd} let us focus here on the simpler two--flavor
case. Among all the resonances, the $\Delta (1232)$ plays a particular
role for essentially {\it two} reasons. First, the $N\Delta$ mass
splitting is a small number on the chiral scale of 1 GeV,
\begin{equation}
\Delta \equiv m_\Delta - m_N = 293 \, {\rm MeV} \simeq 3 F_\pi \,\, ,
\end{equation}
and second, the couplings of the $N\Delta$ system to pions and photons
are very strong,
\begin{equation}
g_{N\Delta \pi}  \simeq 2 g_{NN\pi} \,\, .
\end{equation}
So one could consider $\Delta$ as a small parameter. It is, however,
important to stress that in the chiral limit, $\Delta$ stays finite
(like $F_\pi$ and unlike $M_\pi$). Inclusion of the spin--3/2 fields
like the $\Delta (1232)$ is therefore based on phenomenological
grounds but also supported by large--$N_c$ arguments since in that
limit a mass degeneracy of the spin--1/2 and spin--3/2 ground state
particles appears. Recently, Hemmert, Holstein and Kambor\cite{HHK}
proposed a systematic way of including the $\Delta (1232)$ based on an 
effective Lagrangian of the type ${\cal L}_{\rm eff} [U, N , \Delta ]$
which has a systematic ``small scale expansion'' in terms of {\it
  three} small parameters (collectively denoted as $\epsilon$),
\begin{equation}
\frac{E_\pi}{ \Lambda} \,\, , \quad \frac{M_\pi}{\Lambda} \,\, , \quad
\frac{ \Delta}{\Lambda} \,\, ,
\end{equation}
with $\Lambda \in [ M_\rho, m_N, 4\pi F_\pi ]$. Starting from the
relativistic pion--nucleon-$\Delta$ Lagrangian, one writes the
nucleon ($N$) and the Rarita--Schwinger  ($\Psi_\mu$)
fields  in terms of velocity eigenstates (the nucleon four--momentum
is $p_\mu = m v_\mu + l_\mu$, with $l_\mu$ a small off--shell 
momentum, $v \cdot l \ll m$ and similarly for the $\Delta (1232)$\cite{jm}),
\begin{equation}
N = {\rm e}^{-imv \cdot x} \, (H_v + h_v) \, , \,\,\,
\Psi_\mu = {\rm e}^{-imv \cdot x} \, (T_{\mu \,v} + t_{\mu \,v}) \, ,
\end{equation}
and integrates out the ``small'' components $h_v$ and $ t_{\mu \,v}$
by means of the path integral formalism developed in Ref.\cite{bkkm}
The corresponding heavy baryon effective field theory 
in this formalism does not only
have a consistent power counting but also $1/m$ suppressed vertices
with fixed coefficients that are generated correctly (which is much simpler
than starting directly with the ``large'' components and fixing these
coefficients via reparametrization invariance). Since the spin--3/2
field is heavier than the nucleon, the residual mass difference 
$\Delta$ remains in the spin--3/2 propagator and one therefore has to
expand in powers of it to achieve a consistent chiral power counting.
The technical details how to do that, in particular how to separate
the spin--1/2 components from the spin--3/2 field, are given in 
Ref.\cite{HHK}

\section{A few examples}

In this paragraph, I briefly discuss two observables which have been
calculated in the extension of CHPT including the $\Delta (1232)$. 

The first one is related to the
scalar nucleon form factor.\cite{bkmzm} In fact, this calculation
predates the formalism developed by Hemmert et al., i.e. the 
$\Delta (1232)$ is treated as a heavy spin--3/2 field in the 
formalism proposed by Jenkins and Manohar.\cite{jmd} 
Nevertheless, the pertinent results have been expanded in powers of 
$\Delta$ and thus the essence of the small scale expansion is
captured. Consider the pion--nucleon $\sigma$--term,
\begin{equation}
\sigma_{\pi N} (t) = \frac{1}{2} (m_u+m_d) \, \langle p' \, | \bar{u}u
+ \bar{d}d \,| p  \, \rangle \,\, , \quad t = (p' -p)^2 \,\, .
\end{equation}
Of particular interest in the analysis of $\sigma_{\pi N}$ is the
Cheng--Dashen point, $t = 2M_\pi^2 \, , \, \nu=0$ (at this unphysical
kinematics, higher order corrections in the pion mass  are the
smallest) and one evaluates the scalar form factor
\begin{equation}
\Delta \sigma_{\pi N} \equiv \sigma_{\pi N} (2M_\pi^2) - \sigma_{\pi
 N} (0) \,\,\, .
\end{equation}
To one loop and order ${\cal O}(\epsilon^3)$, $\Delta \sigma_{\pi N}$
is free of counter terms and just given by simple 
one loop diagrams,\cite{bkmzm}
\begin{eqnarray}
\label{CDres}  
\Delta \sigma_{\pi N} &=& \frac{3 g_A^2 M_\pi^2}{64 \pi^2 F_\pi^2} \,
\biggl\{ \pi M_\pi + (\pi-4)\Delta - 4 \sqrt{\Delta^2 -M_\pi^2} \ln
\biggl(\frac{\Delta}{M_\pi} + \sqrt{ \bigl(\frac{\Delta}{M_\pi}\bigr)^2 -1 }
\biggr)+ \ldots \biggr\} \nonumber \\ 
&=& (7.4 + 7.5) \, {\rm MeV} \simeq 15 \, {\rm MeV} \,\, ,
\end{eqnarray} 
where the first term comes from the intermediate nucleon and scales
as $M_\pi^3$ whereas the other terms come from the loop graph with
the intermediate $\Delta (1232)$ and scale as $\Delta \, M_\pi^2$,
which are therefore both ${\cal O}(\epsilon^3)$. In the chiral
expansion of QCD, however, the first term is ${\cal O}(q^3)$ while the
second is of order $q^4$. This result agrees nicely with the one of
the recent dispersion--theoretical analysis (supplemented by chiral 
symmetry constraints) of Gasser, Leutwyler and Sainio, $\Delta
\sigma_{\pi N} = (15 \pm 1)\,$MeV.\cite{gls} Notice, however, 
that the CHPT result might be strongly affected by higher order effects and 
SU(3) breaking as the ${\cal O}(q^4)$ analysis of the baryon masses
and $\sigma$--terms presented in Ref.\cite{bora} indicates.
For example, the chiral expansion of the $\sigma$--term carried out
to second order in the quark masses and for three flavors is\cite{bora}
\begin{equation} 
\sigma_{\pi N} (0) = 58.3 \, ( 1 - 0.56 + 0.33) \, {\rm MeV}
= 45 \, {\rm MeV} \,\, 
\end{equation}
which shows a moderate convergence but also indicates the 
importance of the terms of ${\cal O}(q^4)$. A calculation
in the formalism of Hemmert et al. to order to ${\cal O}(\epsilon^4)$
could help to clarify the situation concerning the higher order corrections.

The second example is the calculation of the electric dipole amplitude
$E_{0+}$ for $\pi^0$ production off protons at threshold.\cite{hhk2}
This calculation has only been performed to order $\epsilon^3$ so far
and can thus not yet compete with the $q^4$ calculations done in the
chiral expansion. It nevertheless shows some interesting features of
the small scale expansion. The diagrams with intermediate nucleons
contributing to this order have been first worked out in
Ref.\cite{bgkm} and later in the heavy baryon approach in
Ref.\cite{bkkm} In the $\epsilon$ expansion, there are a few more
graphs, but only the Born graph with an intermediate $\Delta (1232)$
contributes at threshold. The contribution of this graph to $E_{0+}$
scales as 
\begin{equation}
E_{0+} \sim V_{N\Delta\gamma} \cdot S_\Delta \, \cdot V_{N\Delta\pi}
\sim \frac{M_\pi^3}{M_\pi + \Delta} \sim M_\pi^2 \sim \epsilon^2 \,\, 
\end{equation}
while in the chiral expansion this term would be of order $M_\pi^3$ in
$E_{0+}$. The complete result to this order takes the form\cite{hhk2}
\begin{equation}
E_{0+}^{\rm thr} = -C \, \mu \, \biggl\{ 1 - \biggl[
\frac{3+\kappa_p}{2} + \biggl( \frac{m}{4 F_\pi}\biggr)^2 - \frac{4
b_1 \tilde{g}_{N\Delta\pi} m}{ 9 g_{\pi N}
F_\pi}\frac{M_\pi}{M_\pi + \Delta} \biggr] \, \mu 
+ {\cal O}(\mu^2) \biggr\} \,\, ,
\end{equation}
with $C = e g_{\pi N}/ (8 \pi m)$, $\mu = m/ M_\pi \sim 1/7$ and
$\tilde{g}_{N\Delta\pi} = 1.5$.
The coupling constant $b_1$ can be extracted from $\Delta N \gamma$ dynamics.
Numerically, this new contribution is small, $\delta E_{0+}^{\rm thr}
\sim -0.2 \cdot 10^{-3}/M_\pi$, as it is expected from the analysis
where the $\Delta (1232)$ is used to estimate the LECs. However, to
this order in the small scale expansion, this expression is
complete. It serves as a good example of the previous statement that
for threshold observables, it is more economical to use the resonances
as frozen degrees of freedom in terms of LECs. The real virtue of
including the $\Delta (1232)$ in the effective Lagrangian comes when
one considers e.g. photoproduction above threshold.

\section{Outlook}

Presently, we are working on evaluating the $E2/M1$ ratio in the 
region of the $\Delta$ pole and the corrections to the P--wave LETs 
for pion photoproduction due to the $\Delta(1232)$ to order 
$\epsilon^3$.\cite{bhkm} Most of the work is already done, but results 
can not yet be reported. It is important to stress how our calculation
differs from the one of Butler et al.\cite{bss}, who found a range
for $E2/M1$, 4\%$\le |E2/M1| \le$9\% and an imaginary part which is of
the same size than extracted from the new MAMI data via speed plot 
techniques.\cite{hdt} First, they work in SU(3) and include the spin--3/2
decuplet without expanding in $m_{3/2} - m_{1/2}$. Second, one counter
term is simply dropped amd $1/m$ corrections are not consistently included.
Third, the remaining parameters are chosen in ranges such that kaon 
loop contributions are effectively suppressed. In the light of the work
presented in\cite{bora} it is certainly necessary to repeat this calculation
for SU(2) taking into account all terms to order $\epsilon^3$. Even when that
is done, one still might have to go one order further. We hope to be able
to report on these results soon. 

\section{Acknowledgements}

I am grateful to V\'eronique Bernard and Norbert Kaiser for allowing me
to present some results prior to publication and to Thomas Hemmert and
Joachim Kambor for setting the stage to do calculations in the
$\Delta$ region. I am also grateful to
Martin Moj\v zi\v s for communicating his results prior to publication. Many
thanks to Harry Lee and Winston Roberts for inviting me and the staff at
the INT for efficient organization. This work was supported in part by
the Deutsche Forschungsgemeinschaft, grant Me864-11/1.

\end{document}